\documentclass{jfm}
\usepackage{graphicx}
\usepackage{natbib}
\usepackage{color}
\usepackage{latexsym}
\usepackage{subfig}
\usepackage{natbib}
\usepackage{upgreek}
\usepackage{mathrsfs}
\usepackage{float}
\usepackage{caption}
\usepackage{soul}
\usepackage{textcomp}
\usepackage{stmaryrd}
\usepackage{amsmath}
\usepackage{mathtools}
\usepackage{epstopdf, epsfig}
\usepackage{amsmath,etoolbox}
\usepackage{bigints}
\usepackage{lipsum}

\usepackage{xargs}                      % Use more than one optional parameter in a new commands
\usepackage[pdftex,dvipsnames]{xcolor}  % Coloured text etc.
\usepackage[colorinlistoftodos,prependcaption,textsize=small]{todonotes}

\AtBeginEnvironment{pmatrix}{\setlength{\arraycolsep}{2pt}}
\def\ee{{\rm e}}
\def\ii{{\rm i}}

\title{Predicting vortex merging and ensuing turbulence characteristics in shear layers from  initial conditions}

\author{Anirban Guha\aff{1},\aff{2}
  \corresp{\email{anirbanguha.ubc@gmail.com}} \and Mona Rahmani\aff{3}
 }

\affiliation{
\aff{1} Institute of Coastal Research, Helmholtz-Zentrum Geesthacht,, Geesthacht 21502, Germany.
\aff{2} School of Science and Engineering, University of Dundee, Dundee DD1 4HN, UK.
\aff{3}Department of Mathematics, The University of British Columbia, 1984 Mathematics Road, Vancouver, British Columbia V6T 1Z2, Canada.
}

\begin{document}

\maketitle

\begin{abstract}
%Unstable shear layers in environmental and industrial flows roll up into a series of vortices, which often form complex nonlinear merging patterns like pairs, triplets, etc. These patterns crucially determine the subsequent turbulence, mixing and tracer transport. Based on classical Kelvin-Helmholtz instability theory, we propose a simple measure that accurately predicts the late-time, highly nonlinear merging patterns solely from the linear initial conditions. The predictions of this measure are substantiated using direct numerical simulations and vortex method. Additionally we show that this measure has significant implications in the ensuing turbulence characteristics. 

Unstable shear layers in environmental and industrial flows roll up into a series of vortices, which often form complex nonlinear merging patterns like pairs and triplets. These patterns crucially determine the subsequent turbulence, mixing and scalar transport.
We show that the late-time, highly nonlinear merging patterns are predictable from the   linearized initial state. The initial asymmetry between consecutive wavelengths of the vertical velocity field provides an effective measure of the strength and pattern of vortex merging.  The predictions of this measure are substantiated using direct numerical simulations. We also show that this measure has significant implications in determining the route to turbulence and the ensuing turbulence characteristics. 

%Based on classical Kelvin-Helmholtz instability theory, we propose a simple measure that accurately predicts the late-time, highly nonlinear merging patterns solely from the linear initial conditions.

% Using linear theory, we show that the initial asymmetry between consecutive wavelengths of the vertical velocity field provides a simple and accurate measure of the late-time, highly nonlinear merging patterns.

% patterns emerge due to an
% initial asymmetry between consecutive wavelengths of the vertical velocity field 

% Using linear theory, we show that the initial asymmetry between consecutive wavelengths of the vertical velocity field provides a simple and accurate measure

% of the late-time, highly nonlinear merging patterns. The predictions of this measure are substantiated using direct numerical simulations and vortex method. Additionally we show that this measure has significant implications in the ensuing turbulence characteristics. 
%We also show evidences of the robustness of vortex merging in the presence of secondary instabilities.  
\end{abstract}

\begin{keywords} 
Vortex merging, Kelvin-Helmholtz instability, turbulence, shear flows
\end{keywords}
\maketitle
\section{Introduction}
\label{sec:Intro}
Vortex merging  is an important mechanism in the evolution process of a series of adjacent  co-rotating vortices \citep{lansky1997theory}. This aesthetically pleasing pattern formation has appealed both artists and scientists, an evidence of the former is the famous painting --\emph{The Starry Night} by 
the 19$^{th}$ century Dutch post-impressionist 
Vincent van Gogh. Vortex merging commonly occur in  shear layers in the oceans \citep{flament2001vortex}, terrestrial \citep{buban2016formation} and planetary atmospheres \citep{mac1986merging}, impinging jets \citep{popiel1991visualization,hwang2001heat},  combustible  jet flames \cite[]{demare2001role}, as well as in less regularly expected areas like the wake of mechanical heart valves \cite[]{bluestein2000vortex}.
The primary vortices arise from the growth of Kelvin-Helmholtz (KH) instabilities on the interface of two fluids with different streamwise velocities. Through the process of merging, two (or more) neighbouring vortices are advected toward each other and merge into a larger vortex \cite[]{zaman1980vortex}.  Vortex merging can significantly enhance the mixing induced by the evolution and turbulent breakdown of KH billows by providing more stirring of the fluids in the two fluid layers \cite[]{rahmani2014effect}. This enhanced mixing has important implications for the estimations of vertical mixing of heat, nutrients and pollutants in the atmosphere, oceans and lakes \cite[]{ivey2008density}. %Vortex merging in the wake of mechanical heart valves strongly affects the dispersion process of blood platelets \cite[]{bluestein2000vortex}.
%In turbulent combustion, the ignition distance, as well as the stability and structure of the jet flames are influenced by the pairing of roll-ups  emerging from the KH instabilities on the jet \cite[]{demare2001role}.
%Vortex pairing has been shown to play a key role in heat transfer and flow structure in impinging jets \cite[]{hwang2001heat,popiel1991visualization}.

\begin{figure}
\centering
\includegraphics[width=1.0\linewidth]{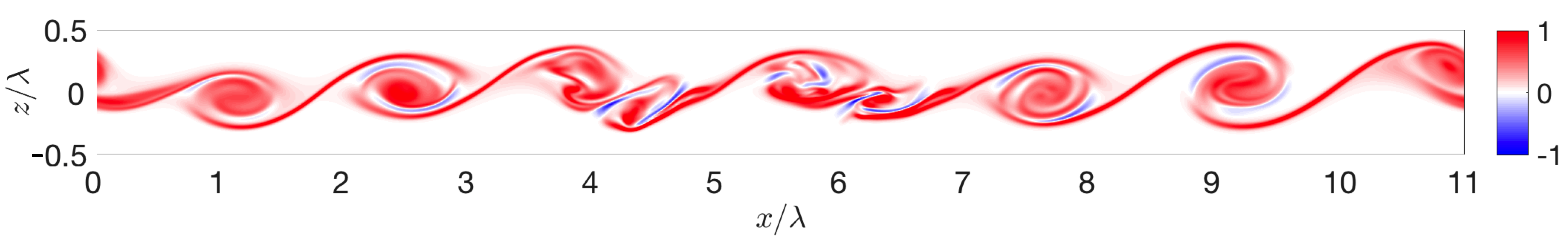}
  \caption{Irregular vortex merging patterns of eleven wavelengths, $\lambda=2\pi/k$, (of the primary KH) in a density stratified shear layer initially perturbed with random noise (results from a fully 3D DNS). {Two pairing events are currently undergoing at the time instant shown, while two pairing events have already taken place. Hence, although initially there were eleven primary KH billows, at present only nine vortices are left. }
  Vorticity contours have been plotted. Negative vorticity primarily ensues from the baroclinic generation.
  }
\label{fig:11L}
\end{figure}

%Another example is the dispersion of particles and different formations of sand dunes due to the effects of pairing of vortices in the wakes behind these dunes \cite[]{muller1986vortex}. 
%that provides a generalized description of vortex merging (pairing, triplets, etc) in shear layers, We show that geometric patterns and strengths of vortex merging are predictable from the linearised initial state, and propose a measure in this regard.
In this paper we show that the complex, \emph{nonlinear} vortex merging patterns  are predictable from the \emph{linear initial conditions}. %and to this end, we propose an accurate measure. 
%report a   that provides an accurate prediction of the complex, \emph{nonlinear} vortex merging patterns in shear layers simply from the \emph{linear initial conditions}. %{ Vortex pairing is the most commonly observed merging process. In shear layers, it occurs due to the growth of the first subharmonic component, i.e. instabilities with twice the wavelength of the primary KH \cite[]{ho1984perturbed}. As an example, the}
Such  merging patterns
%{of KH billows}
are illustrated in figure \ref{fig:11L}, where a numerical simulation of a shear layer susceptible to KH instabilities has been initiated with a pure random noise.
%that does not favour any particular mode.
The asymmetrical structure of the KH billows leads to the merging of neighbouring vortices, albeit in a highly non-uniform way. { Similar  merging patterns have also been observed in  laboratory experiments \citep{ho1982subharmonics},
%: (i) three or four vortices undergoing simultaneous merging, (ii)  a pair of merged vortices merging with a third vortex, and (iii) two pairs of merged vortices agglomerating together.
which indicate the growth of higher subharmonics. The numerical work of \cite{baty2006kelvin}, which investigated multiple wavelengths of the primary KH mode occurring in magnetized jets relevant to astrophysical flows, also reported complex merging patterns. Experiments by \cite{unal1988vortex} and \cite{williamson88} showed the occurrence of vortex pairing as well as evidence of higher subharmonics in the near wake of a circular cylinder. Equivalent experimental observations were also reported by \cite{rajagopalan2005flow}, who further concluded that the near wake region behaves similar to a shear layer. The field observations of the oceanic horizontal shear layer past the island of Hawaii by \cite{flament2001vortex} revealed that the anticyclonic vortices formed on this shear layer proceeded to successively merge to form the first, second and third subharmonics of the original instability. 
%showed that the near-wake of a circular cylinder behaves similar to a shear layer , vortex pairing and higher subharmonic components were observed.
%We note here that subharmonics are known to have important consequences in allied problems, for example, in the wake of a circular cylinder, subharmonics arising in the vortex street lead to a period-doubling route to chaos \citep{sheard2005evolution}.
}

%Quantitative investigations on the physics of multiple vortex merging and turbulent coherent structures are rare in the literature.
%Both numerical simulations and experimental observations show that the initial conditions of the phase difference and amplitude ratio between the primary KH and the subharmonic components significantly influence the late-time evolution of the merging process.
Experiments \citep{husain1995experiments,ho1982subharmonics}, numerical simulations \citep{patnaik1976numerical,corcos1984mixing,dong2018} {and theoretical models \citep{kelly67,nikitopoulos87s}} on vortex pairing have revealed that depending on the phase difference (between the primary KH and the \emph{first} subharmonic component), the outcome in the nonlinear stages can be radically different. An  optimal phase difference leads to a ``rolling interaction" of vortices, while ``shredding interaction" occurs at a non-optimal value of phase difference.
 %Thus the phase difference can lead to radically different outcomes of the 
  The former mode involves a dramatic engulfment of fluid by the merging vortices, whereas in the latter case,  one weaker vortex is merely shredded by the strain field of the stronger vortex. 
  %\textcolor{cyan}{The importance of $\Phi_2$ was further investigated through controlled experiments by \cite{husain1995experiments}, who reported disagreements between experimental results and the predictions of weakly nonlinear theory.}
%These two distinct modes were later identified in other numerical studies \cite[]{corcos1984mixing,riley1980direct}.
%{ While theoretical analysis suggests that  suppression of vortex pairing occurs only at a very specific phase difference and is not realized at any other phase difference \citep{monkewitz1988subharmonic}, shear layer experiments show contradictions and finds suppression over a wide range near the non-optimal phase difference angle \citep{hajj1993fundamental,husain1995experiments}.}
%It has been demonstrated that self-sustained vortex pairing in a jet is the
%manifestation of a subharmonic linear Floquet instability of the underlying periodic
%vortex street.
The sensitivity of  vortex pairing to  phase difference has also been established in  laboratory experiments with round jets \cite[]{arbey84,broze1994nonlinear,paschereit1995experimental}. This phase difference  can be used as a parameter for controlling the transitional behaviour, turbulence structure and mixing in shear layers \cite[]{hajj1993fundamental,dong2018} and round jets \cite[]{cho1998vortex}, and even in noise control applications \cite[]{bridges1987roles,schram2005theoretical}. 

{While pairing is a nonlinear process and its dynamics has traditionally been described  using weakly nonlinear theory \citep{kelly67,monkewitz1988subharmonic}, a recent  study \citep{shaabani2019vortex}  has demonstrated that self-sustained vortex pairing in a time-periodic jet is the
manifestation of a subharmonic linear Floquet instability of the underlying time-periodic vortex street. The present work also reveals that vortex pairing (and other merging patterns like triplets, quadruplets, etc.)  in steady shear layers arise from a linear mechanism.
However we take a very different approach in the present work -- we provide a simple and mechanistic understanding of the role of subharmonics (not only the first, but any arbitrary subharmonic mode) on vortex merging in shear layers, which allows us to predict late-time merging patterns in multiple wavelengths of KH waves simply from the linear initial state. }

\section{Problem layout}
 We assume the classic vortex-sheet profile of KH instability \citep{batchelor2000introduction,drazin2004hydrodynamic}:
 %$\bar{u}=U \mathrm{sgn}(z)$ {and $\bar{\rho}=0.5(\rho_1+\rho_2)+0.5(\rho_1-\rho_2)\mathrm{sgn}(z)$},
%The basic mathematical and physical understanding of KH  is given  in classic texts like  \cite{batchelor2000introduction} and \cite{drazin2004hydrodynamic}. The following background profile is assumed: 
%\begin{subequations}
\renewcommand{\theequation}{\arabic{section}.\arabic{equation}a,b}
\begin{equation}
\bar{u}\left(z\right)=\begin{cases}
U & z>0\\
-U & z<0
\end{cases}\,\,\,\, \mathrm{and}\,\,\,\,\bar{\rho}\left(z\right)=\begin{cases}
\rho_1 & z>0\\
\rho_2 & z<0,
\end{cases}
\label{eq:KH_base}
\end{equation}
\renewcommand{\theequation}{\arabic{section}.\arabic{equation}}
where
 $\bar{u}$ and $\bar{\rho}$ respectively denote the streamwise ($x$) background velocity and {background density (with $\rho_2\geq \rho_1$)}, and $z$ points vertically upwards.  The vortex sheet  is perturbed by an infinitesimal sinusoidal disturbance of the form $\exp\left[ \ii\left( kx/\beta + ly -\Omega_\beta t\right)\right] $, where  $k/\beta\in \mathbb{R}^+$ and $l \in \mathbb{R}^+$ are respectively the streamwise and spanwise wavenumbers, and  $\Omega_\beta \in \mathbb{C}$ is the frequency.
 \renewcommand{\theequation}{\arabic{equation}}

%\noindent where  $\big(u_\beta(x,y,z,t),\,v_\beta(x,y,z,t),\,w_\beta(x,y,z,t)\big)$ and $\eta_{_\beta}(x,y,t) =\hat{\eta}_\beta\exp[ \ii( kx/\beta + ly -\Omega_\beta t)]$ respectively denote the perturbation velocity and interface elevation, and $\tilde{k}=\sqrt{(k/\beta)^2+l^2}$.
Although the velocities at the interface are not defined, they can be calculated by taking the arithmetic mean of the velocities just above and below the interface.  
The perturbation vertical velocity at the interface is obtained by exploiting the linearized kinematic condition at the interface with  $\bar{u}=0$:
$$  \frac{\partial \eta_\beta}{\partial t} + \bar{u}\frac{\partial \eta_\beta}{\partial x}=w_\beta=-\ii \Omega_\beta\eta_\beta, $$
%The perturbation vertical velocity at the interface is found from the linearized kinematic condition with  $\bar{u}=0$, and  is given by  $w_\beta(x,y,0,t)=-\ii \Omega_\beta\eta_\beta$, 
where $\eta_\beta$ denotes the interface elevation.
% velocity at the interface is found to be
% \begin{equation*}
% w_\beta (x,y,0,t)=-\ii \Omega_\beta\eta_\beta.
% \label{eq:1.3}
% \end{equation*}
 %The perturbation vertical velocity at the interface is obtained by exploiting the linearized kinematic condition at the interface with  $\bar{u}=0$:%
%$ \frac{D\eta_\beta}{Dt} \equiv \frac{\partial \eta_\beta}{\partial t} + \bar{u}\frac{\partial \eta_\beta}{\partial x}=w_\beta=-\ii \Omega_\beta\eta_\beta. $
This shows that $w_\beta$ has the same phase as  $\eta_\beta$ provided $\Omega_\beta$ is purely imaginary (unstable flow with zero phase speed). 

The classic dispersion relation for KH instability under the Boussinesq approximation yields \citet{drazin2004hydrodynamic}:
\renewcommand{\theequation}{\arabic{section}.\arabic{equation}}
\begin{equation}
    \Omega_\beta\equiv \ii \gamma_\beta=\pm \ii \sqrt{\left(U\frac{k}{\beta}\right)^2-At g\tilde{k} },
    \label{eq:disp_KH}
\end{equation}
where $\gamma_\beta$ is the growth rate, $\tilde{k}=\sqrt{(k/\beta)^2+l^2}$, {$At \equiv (\rho_2-\rho_1)/(\rho_1+\rho_2)$ is the Atwood number} and $g$ is gravitational acceleration. Equation (\ref{eq:disp_KH}) shows that for instability, $\Omega_\beta$ is purely imaginary, hence $w_\beta$ and  $\eta_\beta$  have the same phase.
This fact has a very important consequence in vortex merging, which will be elaborated in the following discussions.

%\subsection{2D disturbances}

% Due to the assumed  sinusoidal dependence in $x$, the variables $\eta_\beta$  and $w_\beta$ at the interface are given by  $$\eta_\beta=\hat{\eta}_\beta (t) \ee^{\ii kx/\beta}\,\,\mathrm{and}\,\,w_\beta=\hat{w}_\beta (t) \ee^{\ii kx/\beta}, $$ where `hat' denotes  Fourier amplitude. The variables $\eta_\beta$ and $w_\beta$ are related by the kinematic condition at the interface, linearized form of which reads
% $$\frac{D\eta_\beta}{Dt} \equiv \frac{\partial \eta_\beta}{\partial t} + \bar{u}\frac{\partial \eta_\beta}{\partial x}=w_\beta.$$ Since $\bar{u}=0$ at the interface, substitution of (\ref{eq:1.3}) in the above equation straight-forwardly yields
% \begin{equation}
% \frac{d \hat{\eta}_\beta}{dt}=\frac{k}{\beta}U\hat{\eta}_\beta.
% \label{eq:1.4}
% \end{equation}
%  The above equation shows that the $k/\beta$ mode is unstable, and grows exponentially at the rate $\gamma_\beta\equiv kU/\beta$. 

 For  now we  restrict ourselves to 2D flows ($x$--$z$) with  polychromatic waves in the streamwise direction (note that now $l=0$);  $k$ is the wavenumber of the primary mode and  $\beta$ is its  $(\beta-1)$th subharmonic; $\beta=2,3,\ldots$ Equation (\ref{eq:disp_KH})  shows that the growth rate $\gamma_\beta$ is inversely proportional to $\beta$. Hence the primary mode has the fastest growth rate, and higher the subharmonic, the weaker is its growth rate. 
 %\todo{Anirban, does this mean when At=0, then the growth rate is a linear function of k?}
 We first consider an interface perturbed by a dichromatic disturbance which has modes $k$ and $k/\beta$.
Hence the initial disturbance has the form
\begin{equation*}
\eta(x,0)=\hat{\eta}_{1}(0)\sin\left(kx\right)+\hat{\eta}_{\beta}(0)\sin(kx/\beta+\Phi_{\beta}),
%\label{eq:init_con}
\end{equation*}
where  $\Phi_\beta \in [0,2\pi)$ denotes the phase difference between the primary  and the subharmonic modes. This initial disturbance for a short time would evolve as 
\begin{equation*}
\eta(x,t)=\hat{\eta}_{1}(0)\ee^{\gamma_1 t}\sin\left(kx\right)+\hat{\eta}_{\beta}(0)\ee^{\gamma_\beta t}\sin(kx/\beta+\Phi_{\beta}),
%\label{eq:evol_eqn}
\end{equation*}
provided $t$ is small enough so that linearity of the system is respected. 
%We also note that for 2D situation, 
%$$ \gamma_\beta=\pm \ii \sqrt{\left(U\frac{k}{\beta}\right)^2-A_t g\tilde{k} }$$
%(suffix in $\gamma$ is dropped for the primary mode).
%\subsubsection{Asymmetry between consecutive wavelengths -- a measure of vortex pairing}
\subsection{Quantifying vortex merging from initial asymmetry}
The physical mechanism behind  vortex pairing is explained in figure~\ref{fig:1}. In figure~\ref{fig:1}(a-i), the first subharmonic mode (in red) tends to oppositely displace consecutive wavelengths of the primary mode (in black); the first wavelength (which gives rise to the first KH billow) moves upward while the second (which gives rise to the second KH billow) moves downward.  The horizontal velocity shear advects the center of the billows  toward each other, leading to pairing. If  no asymmetry is introduced by the subharmonic mode, there wouldn't be any pairing; see  figure~\ref{fig:1}(a-ii). The only difference between 
figure \ref{fig:1}(a-i) and figure \ref{fig:1}(a-ii) is the phase-shift, $\Phi_2$. In a very similar manner, the second subharmonic induces pairing in
figure~\ref{fig:1}(a-iii) but does not in figure~\ref{fig:1}(a-iv). The basis of the above argument is the fact that the interface elevation of any given mode and the corresponding vertical velocity are in phase.

Since asymmetry induced by the $w$ velocity in two neighboring wavelengths of the primary mode leads to merging, we propose a quantitative measure that captures this effect.  We consider two consecutive wavelengths of the primary mode and find the integral of the $w$ velocity in each of them. The integrated $w$ velocity due to the primary mode will always cancel out, but that due to the subharmonic mode can remain. We define a quantity $\mathcal{A}$ that measures the  strength of the subharmonic mode induced asymmetry with respect to the $w$ velocity amplitude of the primary mode. Mathematically we write it as follows:
 {
%  \begin{align}
% \mathcal{A}^{(\beta)}_{m,m+1}& =\frac{\hat{w}_{\beta}(0)}{\hat{w}_{1}(0)}\left[\intop_{2(m-1)\pi/k}^{2m\pi/k}\sin\left(\frac{k}{\beta}x+\Phi_{\beta}\right)dx-\intop_{2m\pi/k}^{2(m+1)\pi/k}\sin\left(\frac{k}{\beta}x+\Phi_{\beta}\right)dx\right]\\
% % & =\frac{\hat{\eta}_{\beta}(0)}{\hat{\eta}_{1}(0)}\left|\intop_0^{2\pi/k}\sin\left(\frac{k}{\beta}x+\Phi_{\beta}\right)dx-\int_{2\pi/k}^{4\pi/k}\sin\left(\frac{k}{\beta}x+\Phi_{\beta}\right)dx\right|\nonumber\\
% & =  -\frac{4\beta}{k}\frac{\gamma_\beta \hat{\eta}_{\beta}(0)}{\gamma_1 \hat{\eta}_{1}(0)} \sin^2\left(\frac{\pi}{\beta}\right)\left[\cos\left(\frac{2m\pi}{\beta}+\Phi_{\beta}\right)\right].
% \end{align}
\begin{align}
 \mathcal{A}^{(m,m+1)}_{\beta}(t) &=\frac{\hat{w}_{\beta}(t)}{\hat{w}_{1}(t)}\bigg[\intop_{2(m-1)\pi/k}^{2m\pi/k}\sin\left(\frac{k}{\beta}x+\Phi_{\beta}\right)dx  
  -\intop_{2m\pi/k}^{2(m+1)\pi/k}\sin\left(\frac{k}{\beta}x+\Phi_{\beta}\right)dx\bigg]  \nonumber\\
% & =\frac{\hat{\eta}_{\beta}(0)}{\hat{\eta}_{1}(0)}\left|\intop_0^{2\pi/k}\sin\left(\frac{k}{\beta}x+\Phi_{\beta}\right)dx-\int_{2\pi/k}^{4\pi/k}\sin\left(\frac{k}{\beta}x+\Phi_{\beta}\right)dx\right|\nonumber\\
& =  -\frac{4\beta}{k}\frac{ \gamma_\beta\hat{\eta}_{\beta}(0)}{ \gamma_1\hat{\eta}_{1}(0)} \ee^{(\gamma_\beta-\gamma_1) t}\sin^2\bigg(\frac{\pi}{\beta}\bigg)\cos\bigg(\frac{2m\pi}{\beta}+\Phi_{\beta}\bigg).
\label{eq:asym_main}
\end{align}
 The prefix `$m,m+1$' of $\mathcal{A}_\beta$ signifies that we are comparing the $m$-th wavelength with the $(m+1)$-th wavelength of the primary mode, where the asymmetry is induced by the  $(\beta-1)$-th subharmonic. The asymmetry is predicted after time $t$ based on linear theory. Unless otherwise stated, we will evaluate the asymmetry at the initial time ($t=0$).  $\mathcal{A}^{(m,m+1)}_{\beta}>0$ implies that two nearby vortices  move towards each other and undergo merging, the higher the magnitude, the quicker is the merging.  Likewise, $\mathcal{A}^{(m,m+1)}_{\beta}<0$ implies that the two neighboring vortices move away from each other.
We also note that, although the asymmetry seems directly proportional to $\beta$ in Eq.~(\ref{eq:asym_main}), it is actually the opposite, which can be quickly verified from the terms $\gamma_\beta/\gamma_1$ and $\sin^2\left(\pi/\beta\right)$.

% \begin{figure}
% \centering
% \includegraphics[width=0.5\linewidth]{first_figurepng}
%   \caption{Variation of sample asymmetries with phase-shifts at $t=0$. The  coefficient $(4\beta\gamma_\beta \hat{\eta}_{\beta}(0))/(k\gamma_1\hat{\eta}_{1}(0))$ is assumed to be $1$ in each case. Three consecutive wavelengths of the primary wave is considered. Blue and red color respectively indicate asymmetry induced by the first and the second subharmonics. While solid lines denote asymmetry between the first and the second primary waves, dashed lines denotes the same between the second and the third.}
% \label{fig:pairing_phase}
% \end{figure}

The phases leading to the maximum and minimum magnitudes of $\mathcal{A}^{(m,m+1)}_{\beta}$ are
%\renewcommand{\theequation}{\arabic{section}.\arabic{equation}a,b}
%\begin{subequations*}
\begin{align*}
\Phi_{\beta}^{max}&=m\pi\left(1-\frac{2}{\beta}\right)\,\,\mathrm{and}\,\,\pi\left[1+m\left(1-\frac{2}{\beta}\right)\right],\\
\Phi_{\beta}^{min}&=\pi\left[\frac{1}{2}+m\left(1-\frac{2}{\beta}\right)\right]\,\,\mathrm{and}\,\,\pi\left[\frac{3}{2}+m\left(1-\frac{2}{\beta}\right)\right].
\end{align*}
%\end{subequations*}}
%Figure \ref{fig:pairing_phase} shows the variation of $\mathcal{A}$ with phase-shift. The asymmetry induced by the first and the second subharmonics have been separately computed. 
Figure figure~\ref{fig:1}(b) shows the variation of $\mathcal{A}^{(m,m+1)}_{\beta}$ with phase-shift. The asymmetry induced by the first and the second subharmonics have been separately computed.  An important observation is that asymmetries between two different subharmonics can largely add-up or can largely cancel out. To elaborate this point further, let us assume that the interface is initially perturbed with the primary disturbance and its first and second subharmonics:
\begin{equation}
 \eta(x,0)=\hat{\eta}_{1}(0)\sin\left(kx\right)+\hat{\eta}_{2}(0)\sin\left(\frac{k}{2} x+\Phi_{2}\right)+\hat{\eta}_{3}(0)\sin\left(\frac{k}{3} x+\Phi_{3}\right).
\label{eq:init_con2}
\end{equation}
% \begin{align}
% \eta(x,0)&=\hat{\eta}_{1}(0)\sin\left(kx\right)+\hat{\eta}_{2}(0)\sin\left(\frac{k}{2} x+\Phi_{2}\right) \nonumber\\
% &+\hat{\eta}_{3}(0)\sin\left(\frac{k}{3} x+\Phi_{3}\right).
% \label{eq:init_con2}
% \end{align}
Here 
%$lcm(2,3)=6$ (where `$lcm$' implies least common multiple)
6 consecutive wavelengths of the primary wavelength need to be considered for observing the merging pattern (which would repeat after every six wavelengths). We do not consider any higher subharmonics ($\beta>3$) since the growth-rate of these modes fall rapidly and are therefore of little consequence. 
%As expected,  the asymmetry between the first two wavelengths of the primary wave is not the same as that between the second and the third. 
We define the total asymmetry at $t=0$ between two consecutive wavelengths (of the primary mode) as
\begin{equation}
\mathcal{A}^{(m,m+1)}_{total}= \mathcal{A}^{(m,m+1)}_{2}+\mathcal{A}^{(m,m+1)}_{3}.
\label{eq:asy11}
\end{equation}
%and plot it in figure~\ref{fig:asym_bar} for various initial phases, keeping the amplitude ratios equal and constant.

\begin{figure}
        \centering

 \subfloat[]{\includegraphics[width=0.45\linewidth]{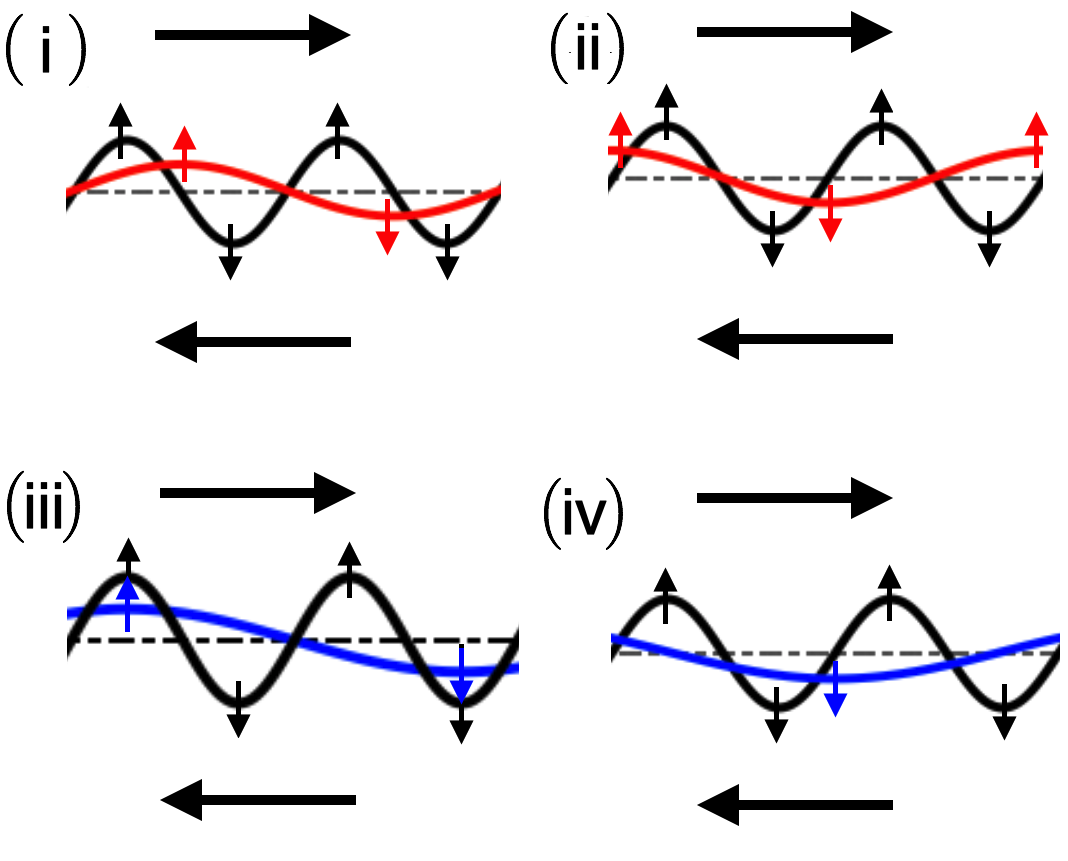}
 }
 \subfloat[]{\includegraphics[width=0.54\linewidth]{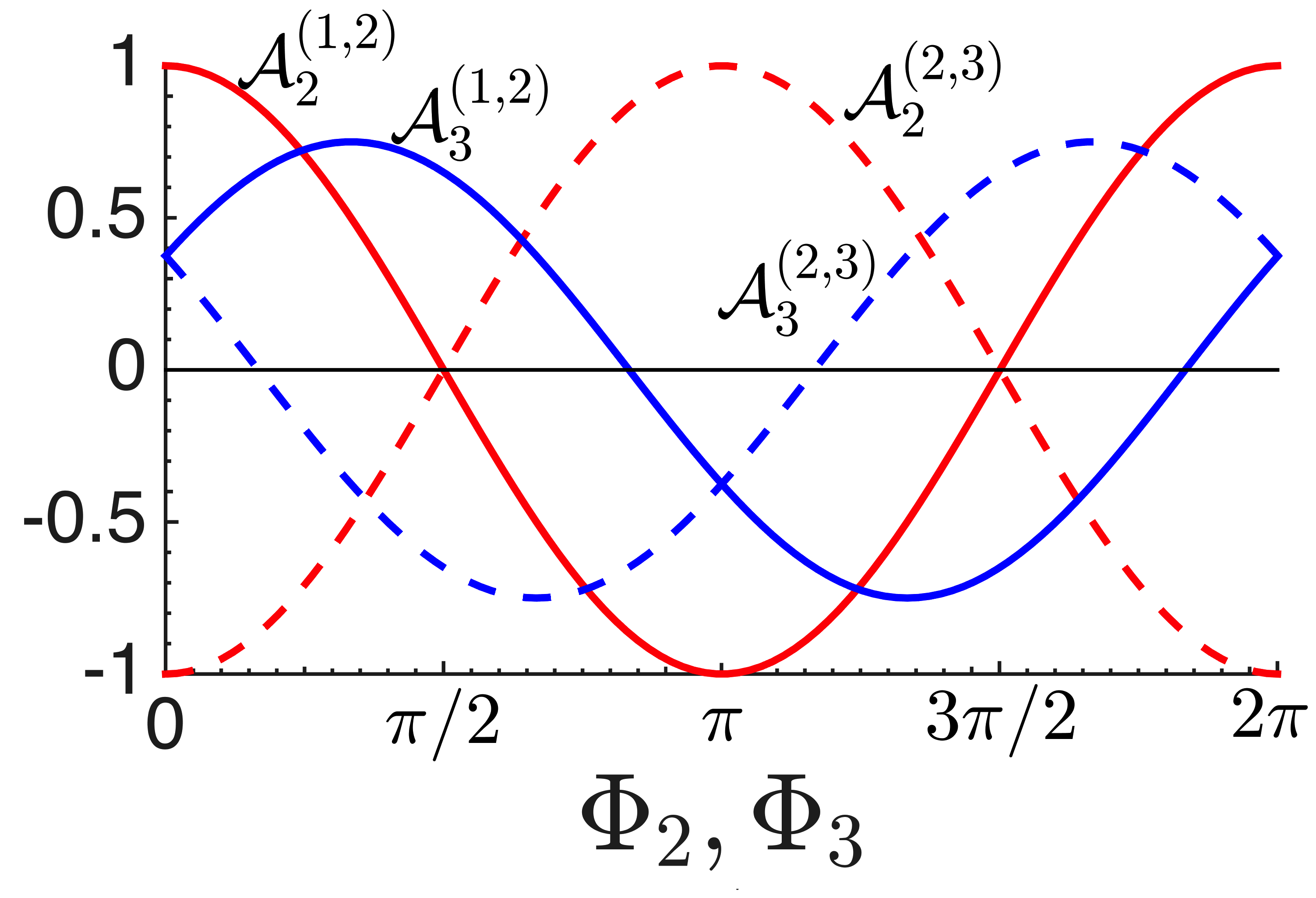}
 }
        \caption{(a) Schematic showing the first two wavelengths of the primary mode of a KH instability, and its first and second subharmonic modes for different phase shifts. The horizontal (vertical) arrows represent horizontal (vertical) velocity. First subharmonic ($\beta=2$) for (i)   $\Phi_2=0$ (maximum asymmetry), and (ii) $\Phi_2=\pi/2$ (minimum asymmetry). Second subharmonic ($\beta=3$) for (iii)  $\Phi_3=\pi/3$ (maximum asymmetry), and (iv)  $\Phi_3=5\pi/6$ (minimum asymmetry). (b) Variation of sample asymmetries with phase-shifts.  Three consecutive wavelengths of the primary wave are considered. Red and blue color respectively indicate asymmetry induced by the first and the second subharmonics. While solid lines denote asymmetry between the first and the second primary waves, dashed lines denotes the same between the second and the third. The coefficient $4\beta \gamma_\beta\hat{\eta}_{\beta}(0)/(k\gamma_1\hat{\eta}_{1}(0))$ of $\mathcal{A}^{(m,m+1)}_{\beta}$ is assumed to be $1$ in each case.
        %(c) second subharmonic ($\beta=3$) for $\Phi_3=0$, and (d) second subharmonic for $\Phi_3=\pi/3$. 
        }
        \label{fig:1}
\end{figure}
\subsection{Predicting vortex merging from initial asymmetry}
Depending on the phases $\Phi_2$ and $\Phi_3$, the asymmetries induced by  these two subharmonics may largely add-up or may largely cancel out.
For simplicity, we restrict to unstratified shear flows ($At=0$).
%which is relevant in horizontal shear setups like atmospheric or oceanic jets or mixing layers with negligible density difference.}  
We span the entire  $\Phi_2$--$\Phi_3$ space to see what combinations of these phase-shifts lead to high or low asymmetry. We also limit our analysis to initial amplitudes satisfying
$$({\hat{\eta}_{2}(0)}/{\hat{\eta}_{1}(0)})^2+({\hat{\eta}_{3}(0)}/{\hat{\eta}_{1}(0)})^2=2.$$ In this respect three cases are considered:
\begin{itemize}
    \item Case C1: $\hat{\eta}_{2}(0)/\hat{\eta}_{1}(0)=1.41$ and $\hat{\eta}_{3}(0)/\hat{\eta}_{1}(0)=0.11$,
    \item Case C2: $\hat{\eta}_{2}(0)/\hat{\eta}_{1}(0)=1$ and $\hat{\eta}_{3}(0)/\hat{\eta}_{1}(0)=1$,
    \item Case C3: $\hat{\eta}_{2}(0)/\hat{\eta}_{1}(0)=0.11$ and $\hat{\eta}_{3}(0)/\hat{\eta}_{1}(0)=1.41$.
\end{itemize}
% (i) C1: $\hat{\eta}_{2}(0)/\hat{\eta}_{1}(0)=1.41$ and $\hat{\eta}_{3}(0)/\hat{\eta}_{1}(0)=0.11$, (ii) C2: $\hat{\eta}_{2}(0)/\hat{\eta}_{1}(0)=1$ and $\hat{\eta}_{3}(0)/\hat{\eta}_{1}(0)=1$, and (iii) C3: $\hat{\eta}_{2}(0)/\hat{\eta}_{1}(0)=0.11$ and $\hat{\eta}_{3}(0)/\hat{\eta}_{1}(0)=1.41$.

 Furthermore, since asymmetry depends on the wavelength-pair `$m,m+1$', we consider the maximum absolute asymmetry, i.e. $\max \left( \left|A_{total}\right| \right)$ among all `$m,m+1$' pairs, and plot it in figure \ref{fig:asym_phase_span}. We observe that  when the first subharmonic is very strong (figure~\ref{fig:asym_phase_span}(a)),  $\max \left( \left|A_{total}\right| \right)$ is nearly independent of $\Phi_3$; the asymmetry is nearly determined by $\Phi_2$.
However, increasing the relative strength of the second subharmonic increases the uniformity of $\max \left( \left|A_{total}\right| \right)$ in the $\Phi_2$--$\Phi_3$ space, as is clearly observed in figure \ref{fig:asym_phase_span}(c). Hence a strong second subharmonic always yields moderate asymmetry independent of the phases.

%\begin{figure}
%\centering
%\includegraphics[width=1.01\linewidth]{mult.png}
%  \caption{(a)-(c): Total asymmetry between consecutive wavelengths (6 such pairs) in an unstratified KH instability.  (d)-(f): Nonlinear evolution of KH instability using vortex method. (g)-(i) Nonlinear evolution of KH instability using DNS.
%  (a,d,g) Case S1, (b,e,h) Case S2, and (c,f,i) Case S3.
%  }
%\label{fig:asym_bar}
%\end{figure}

While the above analysis  revealed  the importance of initial phase differences and amplitude ratios in deciding the `global' picture of asymmetry (using $\max \left( \left|A_{total}\right| \right)$), the `local' asymmetry $\mathcal{A}^{(m,m+1)}_{total}$ (obtained from \eqref{eq:asy11}) is the key in providing quantitative estimates of the propensity towards vortex merging.
To this end  we keep $$\hat{\eta}_{2}(0)/\hat{\eta}_{1}(0)=1\,\, \mathrm{and}\,\, \hat{\eta}_{3}(0)/\hat{\eta}_{1}(0)=1,$$ and vary the initial phase differences:
\begin{itemize}
    \item Case S1: $\Phi_2=0$ and $\Phi_3=\pi$,
    \item Case S2: $\Phi_2=\pi/2$ and $\Phi_3=0$,
    \item Case S3: $\Phi_2=\pi/2$ and $\Phi_3=\pi/2$.
\end{itemize}
%(i)  $\Phi_2=0$ and $\Phi_3=\pi$, (ii)  $\Phi_2=\pi/2$ and $\Phi_3=0$, and (iii) $\Phi_2=\pi/2$ and $\Phi_3=\pi/2$. 
%These cases will be hereafter labelled as S1, S2 and S3.
The value of $\max \left( \left|A_{total}\right| \right)$ is $7$ for S1, which is more than twice of the values corresponding to S2 ($\max \left( \left|A_{total}\right| \right)=3$) and S3 ($\max \left( \left|A_{total}\right| \right)=2.6$).
Figure \ref{fig:asym_bar}(a) corresponds to S1, and shows a moderately positive $\mathcal{A}^{(1,2)}_{total}$ and $\mathcal{A}^{(5,6)}_{total}$, a strongly positive $\mathcal{A}^{(3,4)}_{total}$,  a strongly negative $\mathcal{A}^{(2,3)}_{total}$ and $\mathcal{A}^{(4,5)}_{total}$, and a weakly negative
$\mathcal{A}^{(6,1)}_{total}$. This implies vortices $1$--$2$ and $5$--$6$ will slowly pair, while the pairing of $3$--$4$ would be far rapid. Negative asymmetry  implies that vortices will move away from each other, which we expect for both $2$--$3$ and $4$--$5$. While figure \ref{fig:asym_bar}(a) has alternating positive and negative asymmetries (meaning the maximum extent of merging is \emph{pairing}), figure \ref{fig:asym_bar}(b) reveals a different picture -- both $1$--$2$ and $2$--$3$ have positive asymmetries, while $3$--$4$ has a negative one. This indicates the formation of \emph{triplets}.  Figure \ref{fig:asym_bar}(c) indicates zero asymmetry for vortices $3$--$4$ and $6$--$1$, and are preceded by a negative asymmetry. This indicates that vortices $3$ and $6$ wouldn't undergo pairing. 
%Note that in figure~\ref{fig:asym_phase_span}(b), S2 and S3 have the same $\max \left( \left|A_{total}\right| \right)$, also evident by the eabsolutmaximum asymmetry     
%{We reemphasize here that all these predictions are made \emph{solely} based on initial amplitude ratios and phase shifts.}
The veracity of all these predictions, made \emph{solely} based on initial amplitude ratios and phase shifts, need to be substantiated using accurate numerical simulations. In fact, further late-time predictions from the initial asymmetry is also possible, and would be discussed in the following section.
%that can accurately capture the complex, nonlinear, merging dynamics.  
\begin{figure}
\centering
\includegraphics[width=1.0\linewidth]{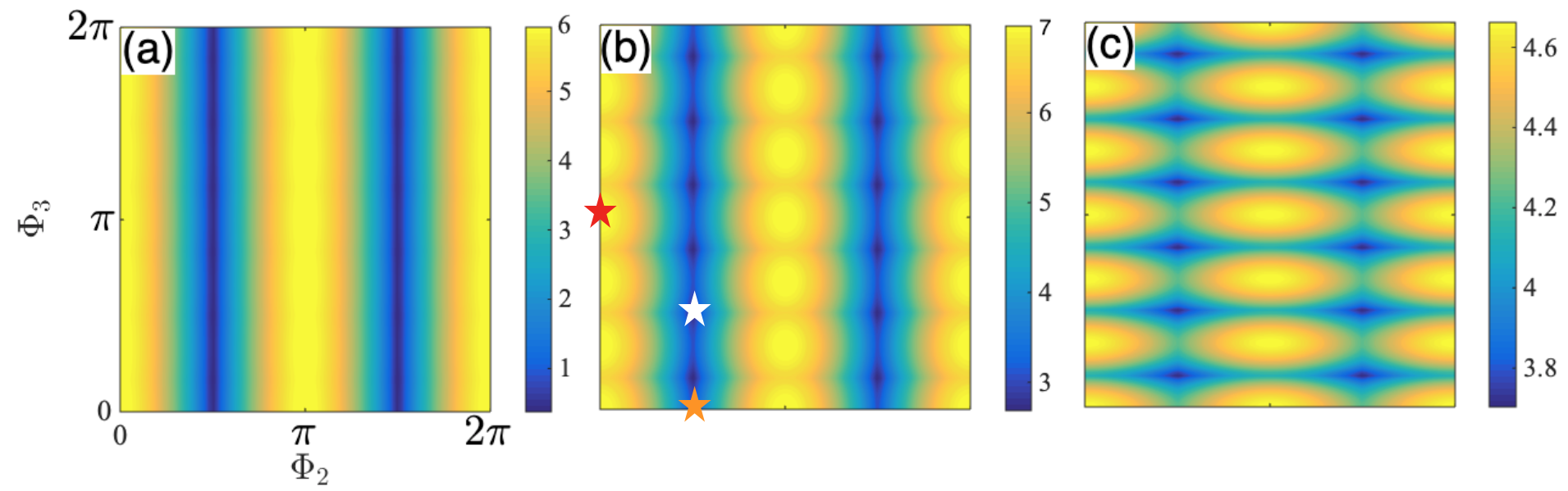}
  \caption{{ Maximum absolute asymmetry  $\max \left( \left|A_{total}\right| \right)$ out of 6 consecutive wavelength pairs in an unstratified KH instability. The entire initial phase-shift space ($\Phi_2$--$\Phi_3$) is spanned. Results are shown for the following amplitude ratios: 
  (a) Case C1, (b) Case C2, and (c) Case C3. Each sub-figure has its own colorbar. Red, orange and white stars respectively denote cases S1, S2 and S3.}}
\label{fig:asym_phase_span}
\end{figure}

\begin{figure*}
\centering
\includegraphics[width=1.0\linewidth]{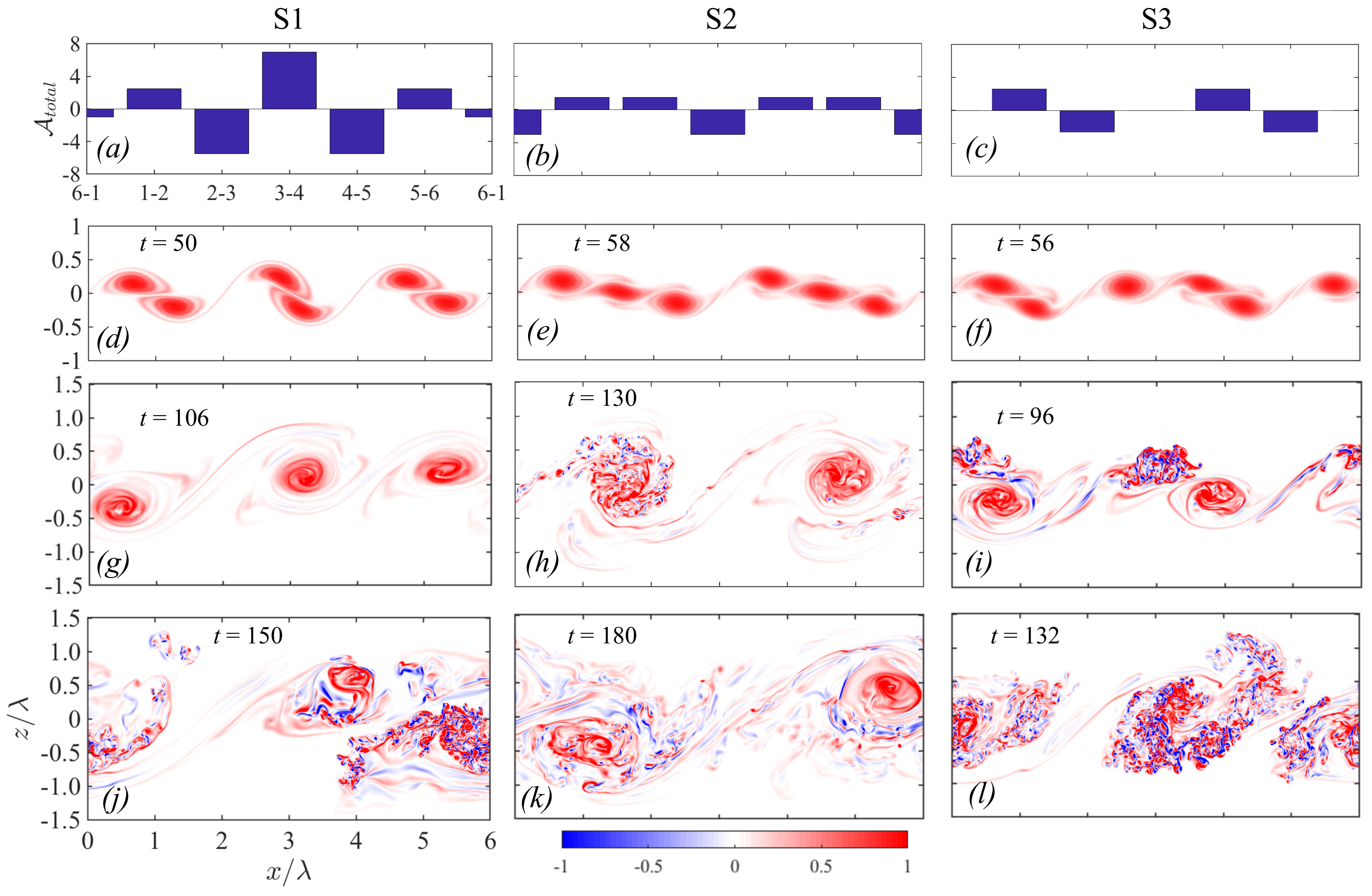}
  \caption{(a)-(c): Total asymmetry between consecutive wavelengths in an unstratified KH instability, inferred only from the linearized initial state. (d)-(f) Nonlinear evolution of KH instability using DNS during the first merging event. (g)-(i) Transition to turbulence in merged KH billows before the second merging event. (j)-(l) Turbulent phase of KH billows just before the last merging event.
  (a,d,g,j) Case S1, (b,e,h,k) Case S2, and (c,f,i,l) Case S3.
The time $t$ is non-dimensionalized by $h_0/\Delta U$.   }
\label{fig:asym_bar}
\end{figure*}

%Coherent structures appearing in high Reynolds number ($Re \equiv \Delta U h_0/\nu$, where $\nu$ is the kinematic viscosity, $\Delta U$ and $h_0$ are respectively the shear velocity and length scales) flows can be effectively approximated using vortex sheets \citep{krasny1986desingularization}. To this end, we implement the 2D vortex sheet method -- a Lagrangian technique where the (desingularized) Birkhoff-Rott equation is solved by discretizing the vortex sheet as an array of point vortices \citep{krasny1986desingularization}, to simulate an initial perturbation given by Eq.\, (\ref{eq:init_con2}). 
%The roll-up and merging patterns for the Cases S1--S3 have been respectively shown in figures\, \ref{fig:asym_bar}(d)-\ref{fig:asym_bar}(f). We find that the observed merging patterns are exactly as predicted by $\mathcal{A}^{(m,m+1)}_{total}$, see figures\, \ref{fig:asym_bar}(a)-\ref{fig:asym_bar}(c).
\section{Numerical simulations of shear instabilities and turbulence}
\label{Sec:3}
%\todo{include second merging from asymmetry}
 Coherent structures appearing in transitional and turbulent shear flows can be accurately simulated using direct numerical simulations (DNS). To this end, we perform simulations using modest Reynolds number ($Re=2000$, where $Re \equiv \Delta U h_0/\nu$; $\nu$ is the kinematic viscosity, $\Delta U$ and $h_0$ are respectively the shear velocity and length scales) DNS, that are more realizable in laboratory and/or industrial settings. %{We solve the 3D incompressible, non-dimensional Navier-Stokes equations. 
%\begin{equation}
%\nabla\cdot\boldsymbol{u} = 0, \;
% D\boldsymbol{u}/Dt = -\rho^{-1}\nabla p + Re^{-1}\nabla^{2}\boldsymbol{u},
% \end{equation}
% where $D/Dt$ denotes the material derivative, $\boldsymbol{u}$ denotes the velocity vector, $p$ denotes the pressure and $\rho$ denotes the density of the fluid.}
The shear layer is initially assumed to be represented by $\bar{u}=\tanh(z)$; the velocity and the vertical coordinate $z$ are respectively non-dimensionalized by  $\Delta U/2$ and $h_0/2$. The governing 3D incompressible, non-dimensional Navier-Stokes equations are
\renewcommand{\theequation}{\arabic{section}.\arabic{equation}a,b}
\begin{equation}
\nabla\cdot\boldsymbol{u} = 0,  \quad
D\boldsymbol{u}/Dt = -\nabla p + Re^{-1}\nabla^{2}\boldsymbol{u},
\end{equation}
\renewcommand{\theequation}{\arabic{section}.\arabic{equation}}
\noindent where $D/Dt$ denotes the material derivative, $\boldsymbol{u}$ denotes the dimensionless velocity vector and $p$ denotes the dimensionless pressure. These equations are solved using a  pseudo-spectral code described in detail by \cite{Winters04} and \cite{Smyth05}. We keep the respective $x$, $y$ and $z$ dimensions as $(L_x,L_y,L_z)=(6\lambda_{KH},0.8\lambda_{KH},3\lambda_{KH})$, where $\lambda_{KH}$ denotes the wavelength of the most unstable KH mode.
%the domain length $L_x$ is set to 6 wavelengths of the most unstable KH mode. The spanwise width of the domain $Ly$ is equal to $0.8$ of one wavelength of the primary KH instability. The domain height $L_z$ is 3 times one wavelength of primary KH mode.
{Each simulation is initially perturbed with the eigenfunctions of the primary KH and its first and second subharmonics with the right phase difference. The amplitudes of the eigenfunction perturbations are sufficiently small to ensure an initial linear growth of each mode. The eigenfunction perturbations are overlaid with random perturbations (of equal or smaller order of magnitude) to trigger 3D instabilities.} The boundary conditions are periodic in both $x$ and $y$ directions and free slip at $z=0$ and $z=L_z$. The number of mesh points used  are: $(N_x,N_y,N_z)=(1152,160,576)$, which  resolves the Kolmogorov length scale.

The DNS results for the cases S1--S3 during the first merging event are respectively shown in figures\, \ref{fig:asym_bar}(d)-\ref{fig:asym_bar}(f). We find that the observed merging patterns are exactly as predicted by $\mathcal{A}^{(m,m+1)}_{total}$, shown in figures\, \ref{fig:asym_bar}(a)-\ref{fig:asym_bar}(c), i.e. neighbouring vortices pair in the case S1, triplets form in the case S2 and alternative pairing/no-pairing occurs in the case S3. The phase relations between the primary mode and its subharmonics also have significant implications for the nonlinear evolution of the vortices in later stages and their turbulent breakdown, as shown in figures \ref{fig:asym_bar}(g)-\ref{fig:asym_bar}(i) and \ref{fig:asym_bar}(j)-\ref{fig:asym_bar}(l).  In  case S1, the merged vortices 1--2 and 5--6 undergo a second merging event, see figure \ref{fig:asym_bar}(g). Finally, the merged vortices 3--4 coalesce with the  merged vortices 1--2--5--6 to form a single vortex;  this coalescence is underway in figure \ref{fig:asym_bar}(j). The development of small-scale structures is delayed until the last merging event since most of the energy extracted from the background shear is spent on pairing (e.g. see \cite{rahmani2014effect}).

For  case S2, during the first merging event each three vortices form a triplet. The  two merged triplets are fairly symmetric and their centers are far apart, hence they resist pairing for a relatively long time, see figure \ref{fig:asym_bar}(h). Finally they undergo pairing after turbulent-like structures have grown in their cores, see figure \ref{fig:asym_bar}(k). In case S3, the left-out vortices 3 and 6 eventually merge with vortices 4--5 and 1--2, respectively (figure \ref{fig:asym_bar}(i)). This merging event is a ``shredding interaction", and leads to the most vigorous disintegration of the core vortex, see figure \ref{fig:asym_bar}(l).
%The reason is that the single vortices with lower strengths are engulfed in the already merged vortices through a ``shredding interaction" rather than going through a ``rolling interaction".
%We also note that the left-out vortices in figure \ref{fig:asym_bar}(i) have developed small-scale structures more energetically compared to the paired vortices.

Interestingly, $\mathcal{A}^{(m,m+1)}_{total}$ can also provide reasonable predictions of the second merging events. In other words, the large-scale 2D patterns observed in figures \ref{fig:asym_bar}(g)-\ref{fig:asym_bar}(i), which respectively occur at $t=150,\,180$ and $132$, are predictable from the $t=0$ state (recall that $\mathcal{A}^{(m,m+1)}_{total}$ is always obtained from \eqref{eq:asy11}, which is the asymmetry evaluated at $t=0$).  Considering case S1, we already found from figure \ref{fig:asym_bar}(a) that the first merging event leads to three vortex pairs: 1--2, 3--4 and 5--6 (figure \ref{fig:asym_bar}(d)). If each primary KH billow has a circulation $\sim \Gamma$, then after the first merging event, each of the merged vortices 1--2, 3--4 and 5--6 will have a circulation $\sim 2\Gamma$. However $\mathcal{A}^{(2,3)}_{total}=\mathcal{A}^{(4,5)}_{total}=-5.5$, while 
$\mathcal{A}^{(6,1)}_{total}=-1$, implying that the merged vortices 5--6 and 1--2 are expected to be far closer to each other than they are individually with 3--4, hence the second merging would be between 1--2 and 5--6. The propensity towards this merging is observed in  figure \ref{fig:asym_bar}(g), and the merging is underway in figure \ref{fig:asym_bar}(j), exactly as predicted.

{
Similar analyses can be done for the cases S2 and S3. For S2, the first merging event would lead to triplets 1--2--3 and 4--5--6, each with a circulation $\sim 3 \Gamma$. Since $\mathcal{A}^{(6,1)}_{total}=\mathcal{A}^{(3,4)}_{total}=-3$, it implies that the two vortex triplets are far removed from each other, and hence resist second merging for a very long time, see figure \ref{fig:asym_bar}(k). The case S3 is more interesting - the first merging produces vortex pairs  1--2 and 4--5, while vortices 3 and 6 are left out  (figure \ref{fig:asym_bar}(f)). Since $\mathcal{A}^{(2,3)}_{total}=\mathcal{A}^{(5,6)}_{total}=-2.6$ while $\mathcal{A}^{(3,4)}_{total}=\mathcal{A}^{(6,1)}_{total}=0$, it implies that vortex 3 would merge with 4--5, while vortex 6 would merge with 1--2. Again, this prediction is exactly found to be true in  figures \ref{fig:asym_bar}(i) and \ref{fig:asym_bar}(l). }
%This is also evident from figure \ref{fig:asym_bar}(d)

\begin{figure}
\centering
\includegraphics[width=1.0\linewidth]{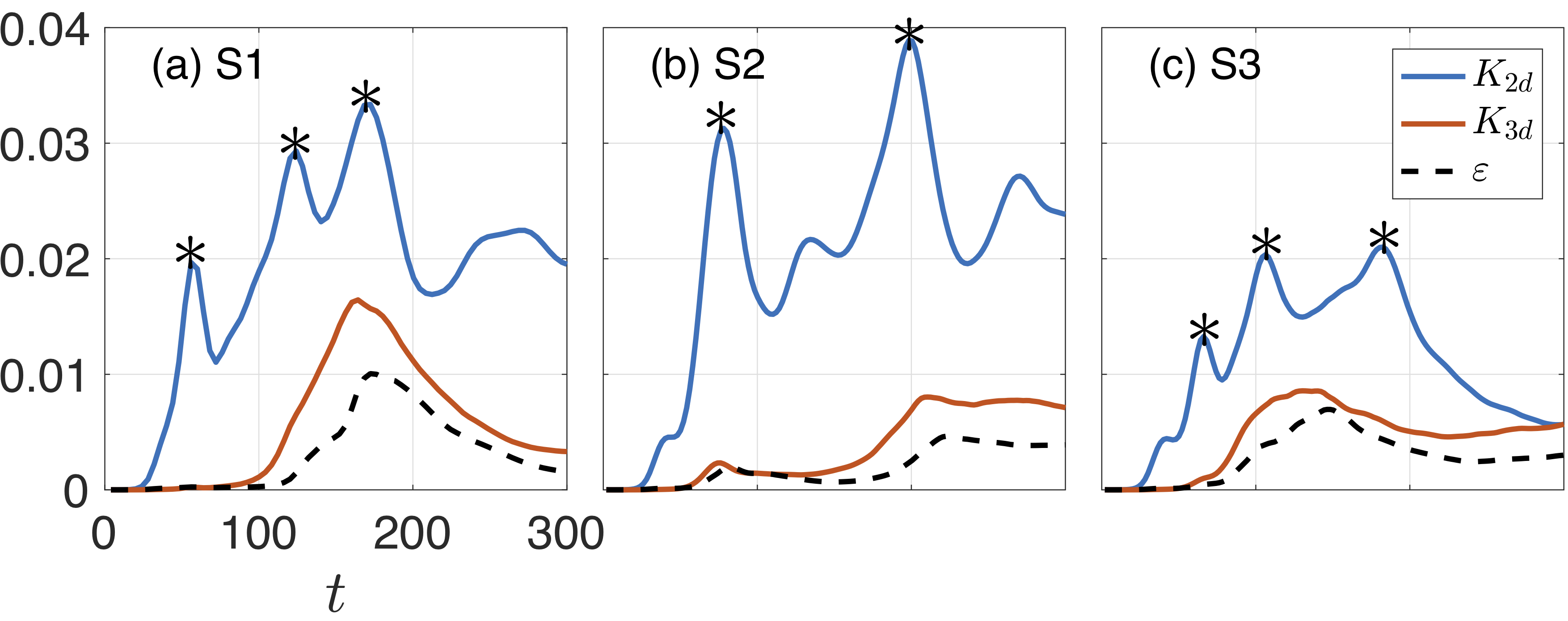}
  \caption{{Time evolution of the dimensionless two- and three-dimensional kinetic energy, $K_{2d}$ and $K_{3d}$ and the rate of viscous dissipation of kinetic energy $\varepsilon$. The parameter $\varepsilon$ has been multiplied by a factor of 20 for plotting. The peaks in $K_{2d}$ corresponding to merging events are marked by stars.}}
\label{fig:ke_eps}
\end{figure}

The growth of turbulent-like coherent structures is a consequence of 3D motions extracting energy from the 2D flow. To quantify the strength of the 2D and 3D motions, and the intensity of turbulence, we utilize the definitions of the 2D kinetic energy: $K_{2d}= \langle\boldsymbol{u}_{2d}\cdot{\boldsymbol{u}}_{2d}\rangle_{xz}$, the 3D kinetic energy: $K_{3d}=\langle\boldsymbol{u}_{3d}\cdot{\boldsymbol{u}}_{3d}\rangle_{xyz}$, and the rate of viscous dissipation of the total kinetic energy: $\varepsilon=Re^{-1}\langle{(\partial u_i / \partial x_j)^2 \rangle}_{xyz}$, with $\langle{\, \rangle}$ denoting the averages in the specified directions  \citep{caulfield2000}. In these definitions the velocity field has been partitioned into three parts: 
%the background 1D velocity, the 2D velocity obtained by averaging the velocity field in the spanwise direction, and the 3D velocity that is the remaining part: 
$\overline{\boldsymbol{u}}(z)={\langle\boldsymbol{u}\rangle}_{xy}$, $\boldsymbol{u}_{2d}(x,z) = {\langle\boldsymbol{u}\rangle}_y - 
{\langle\boldsymbol{u}\rangle}_{xy}$, and 
$\boldsymbol{u}_{3d}(x,y,z) = \boldsymbol{u} - \overline{\boldsymbol{u}}- \boldsymbol{u}_{2d}$. The competition between $K_{2d}$ and $K_{3d}$, and the time evolution of $\varepsilon$, are shown in figure \ref{fig:ke_eps} for the cases S1-S3. Note that after the time shown (i.e. $t>300$) all the motions start to decay. The intensity of the 2D and 3D motions and the viscous dissipation rate varies significantly between the different cases. The highest values in $K_{2d}$ correspond to the case S2, where triplets are formed and the vertical extent  of the merged vortices ({i.e. inertial length scale}) is the largest. The major local peaks in $K_{2d}$ mark the merging events. For example, in cases S1 and S3, three merging events occur, while in the case S2, only two merging events occur.  The global maxima in $K_{3d}$ and $\varepsilon$ occur close to the last merging event in the cases S1 and S2, and close to second merging in the case S3, where the second merging involves the agglomeration of the most asymmetric vortices. The peaks in $K_{3d}$ and $\varepsilon$ are the highest for S1 with the highest $ \mathrm{max} \left( \left|A_{total}\right| \right)$. The values of the peaks of $K_{3d}$ are almost the same for the cases S2 and S3 which have close values of $ \mathrm{max} \left( \left|A_{total}\right| \right)$. 
%However, the dissipation rate, $\varepsilon$, is less directly correlated to $\max(|A_{total}|)$ in the cases S2 and S3. 
We therefore conclude that the initial asymmetry of the primary KH wave with respect to its subharmonic modes, given by the global measure $ \mathrm{max} \left( \left|A_{total}\right| \right)$, has significant implications for the intensity of the ensuing turbulence. 
%in the flow that ensues the growth of primary KH instabilities.
For the highest  $ \mathrm{max} \left( \left|A_{total}\right| \right)$, vortex merging occurs either earlier (as in S1 compared to S2) or more energetically (as in S1 compared to S3). In the former case, the flow has more time to develop small-scale 3D turbulent motions before the decay starts (e.g. see  \cite{rahmani2014effect}) and in the latter case, the 3D motions can extract energy more efficiently from the 2D flow during the active turbulent phase. In both these situations, the higher  $ \mathrm{max} \left( \left|A_{total}\right| \right)$ has led to a higher intensity of turbulence. 

\section{Summary}
\label{sec:Summary}
In summary, we investigated the complex, multiple vortex merging patterns and ensuing turbulence characteristics in a shear layer. { We have considered multiple wavelengths of the primary KH mode, and investigated the effect of higher subharmonics and not just the first subharmonic) on the primary KH for understanding the physics of vortex merging. While related previous studies that considered vortex array \citep{unal1988vortex,rajagopalan2005flow,baty2006kelvin,shaabani2019vortex} have emphasized on the role of subharmonics in vortex merging, a simple physical understanding of the underlying mechanism seems to be missing.
Based on the linear theory of  KH instability arising in the classic vortex-sheet profile, we have provided a mechanistic understanding of the role of subharmonics in vortex merging.} We have shown that  the otherwise symmetric vertical velocity field of two neighboring wavelengths of the primary KH is rendered asymmetric by the presence of a subharmonic mode. This asymmetry, which is fully derived from the linear theory, is found to be the key in deciding the local merging patterns and their strengths. Based on the initial asymmetry of the vertical velocity profile,  we have proposed an effective measure of vortex merging. Our analysis reveals that the highly nonlinear, local merging patterns in shear layers are predictable from the linearized initial state. 
Additionally, we show that subsequent merging patterns can also be predicted from the initial asymmetry. In fact, the highest amount of initial global asymmetry is found to yield the highest level of turbulence intensity. { In summary, the main contribution of this paper is twofold -- (i) to provide a simple, linear, mechanistic description of the role of subharmonics in vortex merging, and (ii) to predict nonlinear vortex merging patterns and ensuing turbulence characteristics from the linearized initial state.}

{
In all our DNS studies reported in \S \ref{Sec:3}, the primary KH and the subharmonics are overlaid with random noise of comparable  amplitude, and yet our model  can still accurately predict the merging patterns from linear initial conditions. 
%\todo{Anirban, should be bring back the following sentence? I think it was good to explain why this happens. Also maybe we don't want to make changes to this paragraph in our next submission?}
%This is due to the fact that the superharmonic noise being stable would decay, and the higher subharmonics, due to their low growth rate would be of little consequence.
Only if the random noise, that can source from other concurrent phenomena in some realistic situations, is significantly stronger (and are nonlinear in nature) than the eigenfunction perturbations, we suspect that the predictive nature of the model, which relies on the linearity of the initial state, would break down. }

 Our findings imply that in many realistic situations, the route to turbulence, coherent structures and turbulent characteristics are strongly dependent on the initial disturbances. Therefore in future, it may be possible to engineer flows to achieve optimal turbulence and mixing.

\section*{Acknowledgement}
    A.G. thanks Alexander von Humboldt foundation for funding. The computational resources for this work were provided by Compute Canada. { The authors also thank Prof. Eyal Heifetz of Tel Aviv University,  Associate Editor Prof. John Dabiri, as well as the two anonymous reviewers for insightful comments and suggestions. }

\bibliographystyle{jfm}
\bibliography{jfm-instructions}
\end{document}